# Modeling of the formation of short-chain acids in propane flames

*F. Battin-Leclerc*[*,1], *A.A. Konnov*[2], *J.L. Jaffrezo*[3] *and M. Legrand*[3]

[1]Département de Chimie-Physique des Réactions, CNRS-INPL, Nancy, FRANCE
[2]Vrije Universiteit Brussel, Brussels, BELGIUM
[3]Laboratoire de Glaciologie et Géophysique de l'Environnement, CNRS, St Martin d'Hères, FRANCE

**Abstract**
In order to better understand their potential formation in combustion systems, a detailed kinetic mechanism for the formation of short-chain monocarboxylic acids, formic ($HCOOH$), acetic ($CH_3COOH$), propionic ($C_2H_5COOH$) and propenic ($C_2H_3COOH$)) acids, has been developed. Simulations of lean (equivalence ratios from 0.9 to 0.48) laminar premixed flames of propane stabilized at atmospheric pressure with nitrogen as diluent have been performed. It was found that amounts up to 25 ppm of acetic acid, 15 ppm of formic acid and 1 ppm of $C_3$ acid can be formed for some positions in the flames. Simulations showed that the more abundant $C_3$ acid formed is propenic acid. A quite acceptable agreement has been obtained with the scarce results from the literature concerning oxygenated compounds, including aldehydes ($CH_2O$, $CH_3CHO$) and acids. A reaction pathways analysis demonstrated that each acid is mainly derived from the aldehyde of similar structure.

**Introduction**

It is now recognized that, in addition to sulfuric and nitric acids, monocarboxylic acids (formic and acetic acids, especially) can contribute to the acidity of gas and aqueous phases of the atmosphere in urban, as well as in remote regions. Natural and anthropogenic sources of monocarboxylic acids have been proposed, but the nature of the individual sources and their relative importance are not established yet [1]. While numerous olefin-ozone reactions can produce formic and acetic acids [2], it is of importance to better understand the potential direct emission of monocarboxylic acids from combustion phenomena, from automotive engines especially.

Detailed analyses of gas-phase pollutants emitted from the exhaust of an engine have been published for diesel [3] and gasoline [4][5][6] engines. These papers describe the presence of noticeable amounts of monocarboxylic acids. These acids are mainly formic, acetic and propionic acids; the formation of this last one being strongly linked with the presence of aromatic compounds in the fuel. In a gasoline engine, the emission of acids corresponds to 4 to 27% of the total amount of emitted hydrocarbons and is larger by a factor of 1.3 to 10 than that of aldehydes [4].

Despite that, acids are very minor oxidation products in most laboratory experimental gas-phase apparatuses and there are very few studies showing their formation during combustion and no kinetic model explaining their formation and consumption pathways. A single paper by Zervas [7] proposes analyses of the formation of acids in a combustion apparatus which can easily be modeled, a laminar premixed flame of propane at atmospheric pressure, but without providing all the data needed for modelling. By chance, a very detailed study of the temperatures and species profiles in a laminar premixed flame of propane at atmospheric pressure has been recently published by Biet et al. [8]. As the two studies have been performed for some common equivalence ratios, we attempted here to implement a new kinetic mechanism of the formation of acids in a model able to well reproduce the results of Biet et al. [8] and to use this global model to simulate the relative formation of monocarboxylic acids in the flame of Zervas [7].

**Kinetic model for the formation of acids in a flame of propane**

The mechanism has been written to reproduce the formation and consumption of formic, acetic, propionic and propenic acids under flame conditions; i.e. the low-temperature reactions of peroxy radicals have not been considered.

This new mechanism was based on the kinetic scheme of the combustion of propane recently proposed to model laminar premixed flame of methane doped with allene or propyne [9]. This $C_3$-$C_4$ reaction base, which was described in details in a previous paper [10], was built from a review of the recent literature and is an extension of our previous $C_0$-$C_2$ reaction base [11]. The $C_3$-$C_4$ reaction base includes reactions involving all the $C_3$ no-oxygenated species and acroleïne, as well as some reactions for selected $C_4$ and $C_5$ compounds and the formation of benzene. In this reactions base, pressure-dependent rate constants follow the formalism proposed by Troe [12] and collisional efficiency coefficients have been included.

To extend this model for description of formation and consumption of acids the relevant sub-mechanism was adopted from the kinetic scheme proposed by Konnov

---

[13]. The number of species was increased by 32 and the number of new reactions was about 200. Selected key reactions discussed in the present work are listed in Table 1.

**TABLE I: SELECTED REACTIONS OF FORMATION AND CONSUMPTION OF $C_1$, $C_2$ AND $C_3$ ACIDS**

The rate constants are given ($k = A\, T^n \exp(-E_a/RT)$) in cc, mol, s, cal units.

| Reactions | A | n | $E_a$ | No |
|---|---|---|---|---|
| *reactions of formic acid* | | | | |
|   reactions of HOCO | | | | |
| CO+OH(+M)=HOCO(+M) | 1.20E+07 | 1.83 | -236.0 | (1) |
|     LOW /7.20E+25 -3.85 1550. / | | | | |
|     TROE /0.6 10. 100000./ | | | | |
| HOCO(+M)=H+CO2(+M) | 1.74E+12 | 0.307 | 32930.0 | (2) |
|     LOW/2.29E+26 -3.02 35070.0/ | | | | |
| HOCO+O2=HO2+CO2 | 1.26E+12 | 0.0 | 0.0 | (3) |
| CH3CO+OH=CH3+HOCO | 3.00E+13 | 0.0 | 0.0 | |
|   reactions of HCOO | | | | |
| HCOO+M=H+CO2+M | 8.70E+15 | 0.0 | 14400.0 | (4) |
| HCOO+O2=CO2+HO2 | 1.00E+11 | 0.0 | 0.0 | (5) |
|   reactions of HCOOH | | | | |
| HCOOH+M=CO+H2O+M | 2.10E+14 | 0.0 | 40400.0 | (6) |
| HCOOH+M=CO2+H2+M | 1.50E+16 | 0.0 | 57000.0 | (7) |
| HCOOH+OH=HOCO+H2O | 1.00E+11 | 0.0 | 0.0 | (8) |
| HCOOH+OH=HCOO+H2O | 1.00E+11 | 0.0 | 0.0 | (9) |
| HCOOH+HCOO=HOCO+HCOOH | 1.00E+11 | 0.0 | 0.0 | |
|   reactions of HOCH2O | | | | |
| HCHO+OH=HOCH2O | 3.40E+06 | 1.18 | -400.0 | (10) |
| HOCH2O=HCOOH+H | 1.00E+14 | 0.0 | 14900.0 | (11) |
| HOCH2O+O2=HO2+HCOOH | 2.10E+10 | 0.0 | 0.0 | (12) |
| *reactions of acetic acid* | | | | |
|   reactions of CH3CO2H | | | | |
| CH3CO+OH=CH3CO2H | 1.00E+14 | 0.0 | 0.0 | (13) |
| CH3CO2H=CH4+CO2 | 7.08E+13 | 0.0 | 74600.0 | (14) |
| CH3CO2H=CH2CO+H2O | 4.47E+14 | 0.0 | 79800.0 | (15) |
| CH3CO2H+OH=CH3CO2+H2O | 2.40E+11 | 0.0 | -400.0 | (16) |
|   reactions of CH3CO2 | | | | |
| CH3CO2+M=CH3+CO2+M | 8.70E+15 | 0.0 | 14400.0 | (17) |
| *reactions of propanoic acid* | | | | |
|   reactions of C2H5CO2H | | | | |
| C2H5CO+OH=C2H5CO2H | 1.00E+14 | 0.0 | 0.0 | (18) |
| C2H5CO2H=C2H6+CO2 | 7.08E+13 | 0.0 | 74600.0 | (19) |
| C2H5CO2H+OH=C2H5CO2+H2O | 2.40E+11 | 0.0 | -400.0 | (20) |
|   reactions of C2H5CO2H | | | | |
| C2H5CO2+M=C2H5+CO2+M | 8.70E+15 | 0.0 | 14400.0 | (21) |
| *reactions of propenic acid* | | | | |
|   Reactions C2H3CO2H | | | | |
| CH2CHCO+OH=C2H3CO2H | 1.00E+14 | 0.0 | 0.0 | (22) |
| C2H3CO2H=C2H4+CO2 | 7.08E+13 | 0.0 | 74600.0 | (23) |
| C2H3CO2H+OH=C2H3CO2+H2O | 2.40E+11 | 0.0 | -400.0 | (24) |
|   reactions of C2H5CO2 | | | | |
| C2H3CO2+M=C2H3+CO2+M | 8.70E+15 | 0.0 | 14400.0 | (25) |

Thermodynamic data of the following species: HOCO, HCOO (formyl radical), HCOOH (formic acid), HCCOH, CH2CHOH (vinyl alcohol), CHOCHO, $CH_3CO_2H$ (acetic acid), $C_2H_5CO$ (propionaldehyde radical), C2H5CHO (propionaldehyde), $CH_2CHCO$ (acrolein radical) were taken from the latest database of Burcat [14]. HiTempThermo database (2006) [15] was used to obtain the data for HCOH (hydroxymethylene), $HOCH_2O$, $HOCH_2OO$, $C_2H_2OH$ (HOCH=CH), $H_2CCOH$, $CH_2CH_2OOH$, $CH_3CHOOH$. Thermochemical data for HCOH (hydroxymethylene), $HOCH_2O$, $HOCH_2OO$, $C_2H_2OH$ (HOCH=CH), $H_2CCOH$, $CH_2CH_2OOH$, $CH_3CHOOH$ and other species were estimated using the software THERGAS developed in Nancy [17], which is based on the additivity methods proposed by Benson [18].

Whenever possible, rate constants have been adopted from the literature and reviews, for instance, of Atkinson et al.[16]; 30 rate constants have been thus found. In many cases, however, rate constants were estimated assuming same reactivity of similar species. For example, the reactivity of $CH_2(OH)_2$, $HOCH_2OO$, $HOCH_2OOH$ and $H_2CCOH$ were assumed analogous to $CH_3OH$, $CH_3O_2$, $CH_3O_2H$ and $C_2H_2OH$ respectively. Strictly speaking this approach could be more or less valid for true homologues; remote similarity could be misleading due to resonance stabilization of radicals, significant modifications of the bond dissociation energy, etc. On the other hand this is probably the only way for exploratory modelling of the formation of acids, which are usually not included in the contemporary combustion mechanisms.

A few modifications of the sub-mechanism adopted from the Konnov [13] kinetic scheme were made. The rate constant of reaction (10) of addition of OH to formaldehyde is taken as 0.1% that of the H-abstraction, which is in agreement with the value of 4% proposed by Sivakumaran et al. [19] as an upper limit for this branching ratio at 298 K. This experimental study of the reaction between OH radicals with formaldehyde by laser photolysis leads the authors to think that the rate constant of the addition is very small as no hydrogen atoms could be observed by resonance fluorescence. The rate constants of the combinations between OH and alcoxy radicals was taken as what proposed by Allara et al. [20] for H-atoms. The reactions of propionic and propenic acids have been deduced from those proposed for acetic acid.

**Validation for the modeling of a propane flame**

All the simulations have been performed using PREMIX of CHEMKIN II collection [21] with estimated transport coefficients. Comparisons between experimental results [8] and simulations of a laminar premixed propane flame are shown in fig. 1 for $\phi= 0.9$ and in fig. 2 for $\phi= 0.48$.

As it can be seen in fig. 1a and 2a, a correct agreement is obtained for the consumption of propane and oxygen and the formation of carbon dioxide for both equivalence ratios. Figure 1b and 1c show that the formation of water, hydrogen, carbon monoxide and $C_2$ compounds are also well reproduced by the model at $\phi= 0.9$. Similarly good agreement is obtained for $\phi= 0.48$, as it can be seen in fig. 2b and 2c. Figures 1d and 2d display the profiles of aldehydes showing that the modelling of formaldehyde and acetaldehyde is also acceptable. These figures present also our prediction for propanaldehyde and acroleïne, which are in agreement with the limit of detection of 100 ppm given by Biet et al. [8].

It is worth noting that the decrease in equivalence ratio does not change markedly the formation of carbon monoxide and aldehydes. That is well explained by Biet et al. [8] by defining an "effective equivalence ratio" based on the proportion of oxygen consumed in the flame and showing that it is always close to 1 when oxygen is in excess.

*Figure 1: Modeling of a laminar premixed flame of propane at 1 atm for an equivalence ratio of 0.9. Symbols correspond to experiments [8] and lines to simulations.*

*Symbols correspond to experiments [8] and lines to simulations.*

**Simulation of the formation of short-chain carboxylic acids**

Figure 3 presents the simulated profiles of acids obtained under the same conditions as the flame of Biet et al. [8] using the mechanism which is described above. This figure shows that small amounts of acids are well produced in this flame, the major ones being formic and acetic acids. Amongst $C_3$ acids, the main one is propenic acid and not propanic acid. That is directly related to the bond dissociation energy of C-CO in $CH_2$=CH-CO (26.2 kcal/mol) and in CH3-CH2-CO (11.9 kcal/mol). The unsaturated oxygenated radical is less easily decomposed than the saturated one to give carbon monoxide (by a factor $6 \times 10^{-3}$ at 1400 K), its concentration is then larger, which favours its combination with OH radicals.

This figure also shows that the formation of acids in these lean flames does not change much when decreasing the equivalence ratio. That is in agreement with what was noticed above for the formation of aldehydes, but in disagreement with the experimental results of Zervas [7] showing increase of about a factor 2 of the formation of formic and acetic acids when varying the equivalence ratio from 1.17 to 0.77.

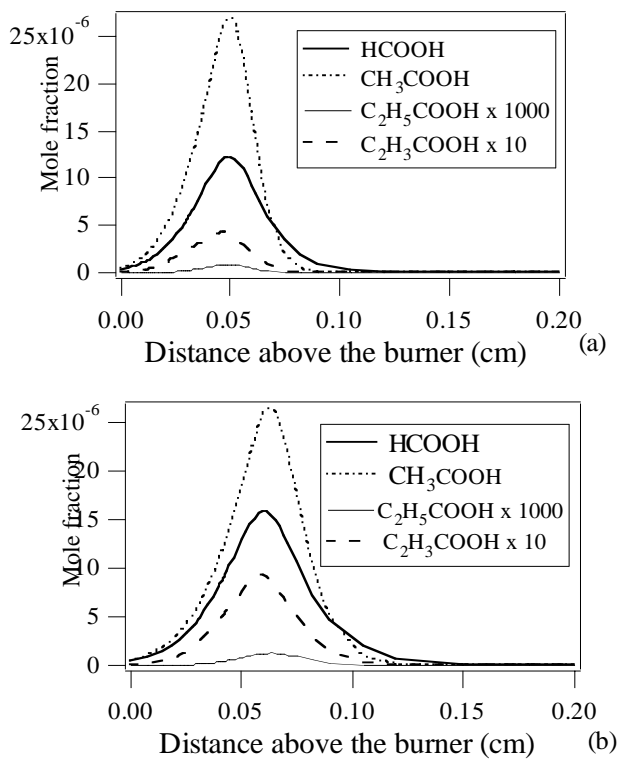

*Figure 3: Modeling of the formation of acids in a laminar premixed flame of propane at 1 atm for an equivalence ratio of (a) 0.9 and (b) 0.48.*

*Figure 2: Modeling of a laminar premixed flame of propane at 1 atm for an equivalence ratio of 0.48.*

Figure 4 displays a comparison between the computed proportions (the ratio between the mole fraction of each acid and the sum of their mole fraction) of acids and the experimental one measured by Zervas et al. [7] for an equivalence ratio of 0.9. In Zervas's work, organic acids were collected in deionised water and analysed by two methods: ionic chromatography/conductimetry detection for formic acid and gas chromatography/flame ionisation detection for other acids. No detail was given about calibration. This figure shows that, apart from the disagreement concerning the nature of the $C_3$ acid, the proportion between $C_1$, $C_2$ and $C_3$ acid is rather well captured by the model.

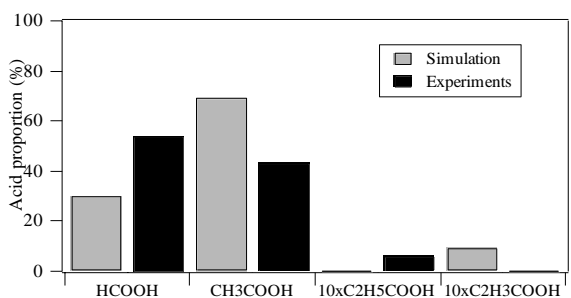

*Figure 4: Comparison between experimental (Zervas [7]) and simulated proportions of acids for an equivalence ratio of 0.9 at the maximum of the peak of formic acid.*

**Analysis of the pathways of formation of acids**

Figure 5 presents the main pathways of formation of the four monocarboxylic acids studied. It can be seen that these acids are mainly derived from the aldehyde having the same structure.

- Formic acid is quasi-exclusively obtained from the addition of OH radicals to formaldehyde, followed by the elimination of a hydrogen atom.
- Acetic acid is mainly formed from the combination of OH and $CH_3CO$ radicals. These last radicals are produced for 40% from reactions of acetaldehyde, but also for 40% from propene and for 20% from vinyl radicals, which are mainly obtained from reactions of ethylene. Propene and vinyl radical can react with O-atoms to give $CH_2CO$ radicals. The reactions of $CH_2CO$ radicals with H-atoms leading to $CH_3CO$ radicals are responsible for 70% of the formation of these radicals.
- Propionic acid is derived from $C_2H_5CO$ radicals by combination with OH radicals. $C_2H_5CO$ radicals are produced for 60% from propanal by H-abstraction by small radicals and for 40% from reactions between methyl radicals and ketene, $CH_2CO$.
- Propenic acid is quasi-exclusively produced from acroleïne by H-abstraction by small radicals, followed by a combination with OH radicals.

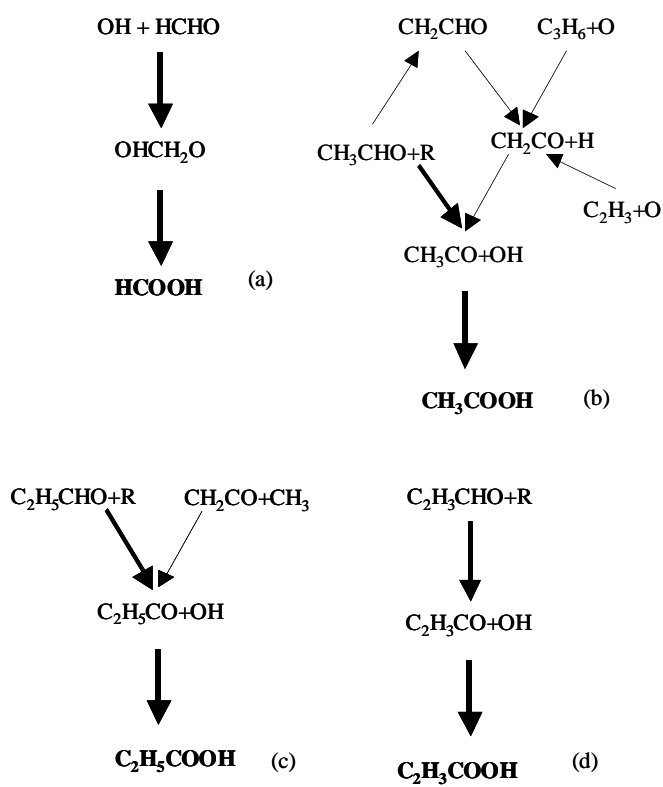

*Figure 5 Main pathways of formation of the four monocarboxylic acids studied under the conditions of figure 3a for a distance above the burner corresponding to the maximum of the peak of formic acid (0.5 cm). The size of the arrows is proportional to the relative flow rates.*

**Conclusion**

A new model for the formation of monocarbocylic acids from $C_1$ to $C_3$ has been proposed and tested to reproduce the formation of these compounds in laminar premixed flames of propane. This model has allowed us to correctly reproduce the proportions between $C_1$, $C_2$ and $C_3$ acids experimentally measured [7], even if the simulations predict the formation of propenic acid while propionic acid has been experimentally reported. Under flame conditions, these acids are mainly derived from the aldehyde having the same structure, by addition of OH radicals for formic acid and by combination of alcoxy and OH radicals for heavier acids.

The predicted amounts of acids are up to hundred times lower than that of aldehydes, explaining why the formation of acid is usually not reported in flame experiments. If the emissions of acids from engines are really larger than that of aldehydes, as reported by Zervas et al. [4] at the exhaust of gasoline engines, low temperature reactions, involving peroxy radicals, should be of greater importance for the formation of acids than those proposed here.